\documentstyle[preprint,aps,epsf]{revtex}

\begin{document}

\title{Renormalization Group Computation of Correlation Functions and Particle
Densities}
\author{Pierre Gosselin$^{a}$ and Herv\'{e} Mohrbach$^{b}$ \\
\textit{
a)Universit\'{e} Grenoble I, Institut Fourier, UMR\ 5582 CNRS-UJF, UFR de
Math\'{e}matiques, BP74, } \and \textit{38402 Saint Martin d'H\`{e}res,
Cedex, France} \and \textit{b) L.P.L.I. Institut\ de\ Physique, 1 blvd
D.Arago, F-57070, Metz, France.}}

\title{Renormalization Group Computation of Correlation Functions and Particle
Densities}
\maketitle

\begin{abstract}
We show that the Renormalization Group formalism allows to compute with
accuracy the zero temperature correlation functions and particle densities
of quantum systems.
\end{abstract}

In \cite{pierre} we showed that the Renormalization Group is an efficient
tool to compute the ground state and first excited energy levels of a one
quantum particle system. The method is the following: instead of computing
directly the effective potential $V_{0}(x_{0})$ defined as a path integral
on the Fourier modes:

\begin{equation}
\exp \left( -{\beta }V_{0}(x_{0})\right) =\int \prod_{1}^{\frac{N}{2}}{\frac{
dx_{m}d\bar{x}_{m}}{{\frac{2\pi \hbar }{\varepsilon \omega _{m}^{2}}}}}\exp
\left( -{\frac{1}{\hbar }}S_{\frac{N}{2}}\right) \,,
\end{equation}
we compute recursively a running potential by integrating each modes after
the other. Note that the Feynman path integral is considered here with a
discretized time $t_{n}={\frac{nT}{N+1}}=n\varepsilon $, with $N$ an
arbitrary large number, and $n=0,\ldots ,N+1$. The Fourier decomposition of
a periodic path $x(t_{n})$ contains only a finite number of Fourier modes 
\begin{equation}
x(t_{n})=x_{0}+{\frac{1}{\sqrt{N+1}}}\sum^{\prime }\exp (i\omega
_{m}t_{n})x_{m}+c.c.,
\end{equation}
where $\sum^{\prime }$ is from $1$ to $\frac{N}{2}$ if $N$ is even and from $
1$ to ${\frac{N-1}{2}}$ if $N$ is odd. The $x_{m}$ are the Fourier modes and 
$\omega _{m}^{2}={\frac{2-cos{\frac{2\pi m}{N+1}}}{\varepsilon ^{2}}}$. The
discrete action is: 
\begin{equation}
S_{\frac{N}{2}}(x)=\varepsilon \sum_{0}^{\frac{N}{2}}{M}\omega
_{m}^{2}|x_{m}|^{2}+\varepsilon \sum_{n=1}^{N+1}V_{\frac{N}{2}}(x(t_{n}))
\end{equation}
In the Local Potential Approximation \cite{pierre},\cite{pierre2}, the
potential at scale m-1 is given by the Wegner-Houghton equation: 
\begin{equation}
V_{m-1}(x_{0})=V_{m}(x_{0})+\frac{1}{\beta }\log (1+\frac{V_{m}^{(2)}(x_{0})
}{\omega _{m}^{2}})+O(\frac{1}{\beta ^{2}})  \label{potfx0}
\end{equation}
In the present work we consider only the zero temperature case and neglect
as a consequence terms of order $\frac{1}{\beta ^{2}}.$ Our experience tells
us that the finite temperature case is more difficult to handle with this
equation \cite{pierre3} and gives less good results than the
Feynman-Kleinert variational method \cite{kleinert}. In order to use this
method to get the correlation functions we have to compute the partition
function in the presence of a source term: 
\begin{equation}
Z\left( j\right) =\int \mathcal{D}xe^{-\int\limits_{0}^{\beta }dt(\frac{1}{2}
\stackrel{.}{x}^{2}+V(x)-j(t)x(t))}  \label{Z}
\end{equation}
Note that the expression is written in the continuum only for convenience.
As in \cite{pierre}, we define recursively the effective potential $
V_{m-1}^{j}(x_{0})$ (which is now $j$ dependent) at scale $m-1$ for a
constant background path $x_{0}=\frac{1}{\beta }\int\limits_{0}^{\beta
}dtx\left( t\right) $, as a path integration on the modes $x_{m}$ and $
x_{-m} $: 
\begin{equation}
e^{-\beta V_{m-1}^{j}(x_{0})}=e^{-\beta V_{m}^{j}(x_{0})}\int \frac{
dx_{m}dx_{-m}}{\frac{2\pi }{\varepsilon \omega _{m}^{2}}}e^{-\varepsilon
\left( \omega _{m}^{2}+V_{m}^{j\left( 2\right) }(x_{0})\right) \left|
x_{m}\right| ^{2}+\left( \varepsilon \sum\limits_{n=1}^{N+1}j(t_{n})\frac{
e^{i\omega _{m}t_{n}}x_{m}}{\sqrt{N+1}}+c.c\right) }  \label{zm}
\end{equation}
At the end of the flow of $V_{m}^{j}(x_{0})$, formula (\ref{Z}) reduces to a
classical partition function with a source term: 
\[
Z\left( j\right) =\int \frac{dx_{0}}{2\pi \beta }e^{-\beta \left(
V_{0}^{j}(x_{0})-j_{0}x_{0}\right) }. 
\]
The gaussian integration in (\ref{zm}) leads to: 
\begin{equation}
V_{m-1}^{j}(x_{0})=V_{m}^{j}(x_{0})+\frac{1}{\beta }\log (1+\frac{
V_{m}^{j(2)}(x_{0})}{\omega _{m}^{2}})-\frac{1}{\beta ^{2}}\frac{
\sum\limits_{n}\varepsilon j(t_{n})\sum\limits_{p}\varepsilon
j(t_{p})e^{i\omega _{m}\left( t_{n}-t_{p}\right) }}{\omega
_{m}^{2}+V_{m}^{j\left( 2\right) }(x_{0})},  \label{rgja}
\end{equation}
or in terms of the Fourier transform of the source: 
\begin{equation}
V_{m-1}^{j}(x_{0})=V_{m}^{j}(x_{0})+\frac{1}{\beta }\log (1+\frac{
V_{m}^{j(2)}(x_{0})}{\omega _{m}^{2}})-\frac{j_{m}j_{-m}}{\omega
_{m}^{2}+V_{m}^{j\left( 2\right) }(x_{0})}.
\end{equation}
By neglecting the source term dependence in the right hand side of equation (
\ref{rgja}), we obtain the following approximate relation: 
\begin{equation}
V_{m-1}^{j}(x_{0})=V_{m}(x_{0})+\frac{1}{\beta }\log (1+\frac{
V_{m}^{(2)}(x_{0})}{\omega _{m}^{2}})-\frac{1}{\beta ^{2}}\frac{
\sum\limits_{n}\varepsilon j(t_{n})\sum\limits_{p}\varepsilon
j(t_{p})e^{i\omega _{m}\left( t_{n}-t_{p}\right) }}{\omega
_{m}^{2}+V_{m}^{\left( 2\right) }(x_{0})}  \label{rgj}
\end{equation}
where now $V_{m}(x_{0})$ is the running potential without source term
defined in equation (\ref{potfx0}). It will be shown below, when computing
the particle density, that this approximation is quite good. Note that the
double sum in the left hand side of (\ref{rgj}) is of order $\beta ^{2}$ so
that the whole source contribution is of order $\beta ^{0}$.

After iteration, (\ref{rgj}) leads to the effective potential $
V_{0}^{j}(x_{0})$ in terms of the effective potential without source $
V_{0}(x_{0})$: 

\begin{equation}
V_{0}^{j}(x_{0})=V_{0}(x_{0})-\frac{1}{\beta ^{2}}\sum_{m=1}^{\frac{N}{2}}
\frac{\int\nolimits_{0}^{\beta }dtj(t)\int\nolimits_{0}^{\beta }dt^{\prime
}j(t^{\prime })e^{i\omega _{m}\left( t-t^{\prime }\right) }}{\omega
_{m}^{2}+V_{m}^{\left( 2\right) }(x_{0})}  \label{potj}
\end{equation}

where for convenience we have replaced the double sum by integrals. It's
then easy to compute the correlation functions using the following relation: 
\begin{equation}
\left\langle x(t_{1})...x(t_{n})\right\rangle =\frac{1}{Z\left( j=0\right) }
\frac{\delta }{\delta j(t_{1})}...\frac{\delta }{\delta j(t_{n})}\int \frac{
dx_{0}}{2\pi \beta }e^{-\beta
V_{0}^{j}(x_{0})+x_{0}\int\nolimits_{0}^{\beta }dtj(t)}  \label{correl}
\end{equation}
Since we consider only the zero temperature limit, the integral in (\ref
{correl}) is dominated by the minimum of the effective potential $V\left( 
\overline{x}_{0}\right) $ so that: 
\begin{equation}
\left\langle x(t_{1})...x(t_{n})\right\rangle =\frac{\delta }{\delta j(t_{1})
}...\frac{\delta }{\delta j(t_{n})}e^{\frac{1}{\beta }\sum\limits_{m=1}^{
\frac{N}{2}}\frac{\int\nolimits_{0}^{\beta }dtj(t)\int\nolimits_{0}^{\beta
}dt^{\prime }j(t^{\prime })e^{i\omega _{m}\left( t-t^{\prime }\right) }}{
\omega _{m}^{2}+V_{m}^{\left( 2\right) }(\overline{x}_{0})}+\overline{x}
_{0}\int\nolimits_{0}^{\beta }dtj(t)}
\end{equation}
As an example the two point correlation function is easily deduced: 
\begin{equation}
\left\langle x(t_{1})x(t_{2})\right\rangle _{j=0}=\overline{x}_{0}^{2}-\frac{
1}{\beta V_{0}^{\left( 2\right) }(\overline{x}_{0})}+\frac{1}{\beta }
\sum_{m=-\frac{N}{2}}^{\frac{N}{2}}\frac{e^{i\omega _{m}\left( t-t^{\prime
}\right) }}{\omega _{m}^{2}+V_{m}^{\left( 2\right) }(\overline{x}_{0})}
\end{equation}

\smallskip

In a similar manner we will now compute the particle density which is
defined as: 
\begin{equation}
\rho (x_{a})=Z^{-1}\int dx\left\langle x\left| e^{-\beta H}\right|
x\right\rangle \delta (x-x_{a})=\frac{1}{\sum_{n}e^{-\beta E_{n}}}
\sum_{n}\left| \psi _{n}(x_{a})\right| ^{2}e^{-\beta E_{n}},
\end{equation}
In the limit $\beta \rightarrow \infty $: 
\begin{equation}
\rho (x_{a})=\left| \psi _{0}(x_{a})\right| ^{2}
\end{equation}
where $\psi _{0}(x_{a})$ is the ground state wave function. In terms of the
path integral the density is: 
\begin{equation}
\rho (x_{a})=\int \mathcal{D}x(t)\delta (x(\tau )-x_{a})\exp \left\{
-\int_{0}^{\beta }dt\left[ \frac{1}{2}\stackrel{.}{x}^{2}+V(x(t))\right]
\right\}
\end{equation}
valuable for any value of $\tau .$ These expression can also be written:

\begin{equation}
\rho (x_{a})=Z^{-1}\int {\frac{dx_{0}}{\sqrt{\frac{2\pi \beta }{M}}}}\int 
\frac{dk}{2\pi }e^{ikx_{a}}\int \mathcal{D}x\delta (\overline{x}
-x_{0})e^{-ikx(\tau )}\exp \left( -S_{\frac{N}{2}}\right)
\end{equation}
Setting $j\left( t\right) =-ik\delta (t-\tau )$ allows to use the previous
formula (\ref{potj}) to directly write:

\begin{equation}
\rho (x_{a})=Z^{-1}\int {\frac{dx_{0}}{\sqrt{\frac{2\pi \beta }{M}}}}\int 
\frac{dk}{2\pi }e^{-k^{2}a^{2}/2+ikx_{a}}e^{-\beta V_{0}(x_{0})}
\end{equation}
where: 
\begin{equation}
a^{2}(x_{0})=\frac{2}{\beta }\sum_{m=1}^{\frac{N}{2}}\frac{1}{{\omega }
_{m}^{2}{+}V_{m}^{(2)}(x_{0})}
\end{equation}
after integrating out $k,$ we obtain: 
\begin{equation}
\rho (x_{a})=Z^{-1}\int {\frac{dx_{0}}{\sqrt{\frac{2\pi \beta }{M}}}}\frac{
e^{-(x_{a}-x_{0})^{2}/2a^{2}(x_{0})}}{\sqrt{2\pi a^{2}(x_{0})}}e^{-\beta
V_{0}(x_{0})}  \label{rho}
\end{equation}
So that in the limit $\beta \rightarrow \infty $: 
\begin{equation}
\rho (x)=\frac{e^{-(x-\overline{x}_{0})^{2}/2a^{2}(\overline{x}_{0})}}{\sqrt{
2\pi a^{2}(\overline{x}_{0})}}
\end{equation}
is a gaussian distribution. The procedure is now very simple. We just have
to determine the flow of the second derivative of the effective potential.
Note that the distribution (\ref{rho}) is the same as the variational
Kleinert's one except that for him $a^{2}(x_{0})=\frac{2}{M\beta }
\sum_{m=0}^{\frac{N}{2}}\frac{1}{{M\omega }_{m}^{2}{+}\Omega (x_{0})}$ where 
$\Omega $ is the variational parameter \cite{kleinert}.

As an example consider the anharmonic oscillator. In \cite{pierre} we showed
that the running potential at scale $m$ is best approximated by a polynomial
of order six: 
\begin{equation}
V_{m}(x)=g_{m}^{0}+\frac{g_{m}^{2}}{2}x^{2}+\frac{g_{m}^{4}}{4!}x^{4}+\frac{
g_{m}^{6}}{6!}x^{6}
\end{equation}
where the initial potential is chosen to be: 
\begin{equation}
V_{\frac{N}{2}}(x)=\frac{1}{2}x_{0}^{2}+\frac{240}{4!}x_{0}^{4}
\end{equation}
Since the minimum of the effective potential is located at $\overline{x}
_{0}=0$ it is then easy to see that $a^{2}(0)=\frac{2}{M\beta }
\sum\limits_{m=1}^{\frac{N}{2}}\frac{1}{{M\omega }_{m}^{2}{+}g_{m}^{2}}.$
From the properties of the correlation function we can also deduce that the
first excited energy level is given by: $E_{1}-E_{0}=\sqrt{g_{0}^{2}}.$

From \cite{pierre} we have: 
\[
\begin{tabular}{|c|c|c|c|c|c|c|c|c|}
\hline
$\lambda $ & $E_{RG}$ & $E_{var}$ & $E_{ex}$ & $E_{1,RG}$ & $E_{1,var}$ & $
E_{1,ex}$ & $a^{2}$ & $a_{var}^{2}$ \\ \hline
$240$ & $1.4982$ & $1.5313$ & $1.50497$ & $5.3368$ & $5.3482$ & $5.3216$ & $
0.1472$ & $0.125$ \\ \hline
\end{tabular}
\]
A comparison between the RG density and the variational density is shown in
fig1. The RG density is very near the exact density given in \cite{kleinert}
. In particular the RG computation greatly improves the value of the density
near $x=0,$ $(\rho _{RG}(0)\approx \rho _{ex}(0)).$

The double well potential case is much more involved. It is well known that
the effective potential at $T=0$ must be a convex quantity. For a very deep
potential the truncation of the running potential is no more a good
approximation of the true effective potential near zero, which is almost
flat. Indeed the truncation of the running potential does not allow a
positive term in the logarithm of the Wegner-Houghton equation and the
approximation breaks down. We have considered the case $\lambda =2.4$ where
an expansion at the order $10$ is necessary to get valuable results: 
\[
\begin{tabular}{|c|c|c|c|c|c|c|c|c|}
\hline
$\lambda $ & $E_{RG}$ & $E_{var}$ & $E_{ex}$ & $E_{1,RG}$ & $E_{1,var}$ & $
E_{1,ex}$ & $a^{2}$ & $a_{var}^{2}$ \\ \hline
$2.4$ & $0.46498$ & $0.549$ & $0.4709$ & $0.8285$ & $1.035$ & $0.7677$ & $
3.48$ & $1.03$ \\ \hline
\end{tabular}
\]
An another problem with our approach is that a convex effective potential is
incompatible with the two picks of the true particle density \cite{kleinert}
(it's clear that the particle will spend most of his time around the two
minima of the potential energy). In statistical mechanics language each spin
fluctuates most of the time around two values (like Ising spin). In fact our
approximation which relates the density to the effective potential breaks
down in this case. Nevertheless, the behavior of our RG density is quite
correct near zero $(\rho _{RG}(0)\approx \rho _{ex}(0))$ and for large
values of $x$. Moreover it is much more better than the variational
computation (fig2.). To get a better result we must go beyond the LPA
approximation, and make the computation for an arbitrary background path
configuration.

\begin{figure}
	\epsfxsize=5cm
	\epsfysize=5cm
	\centerline{\epsffile{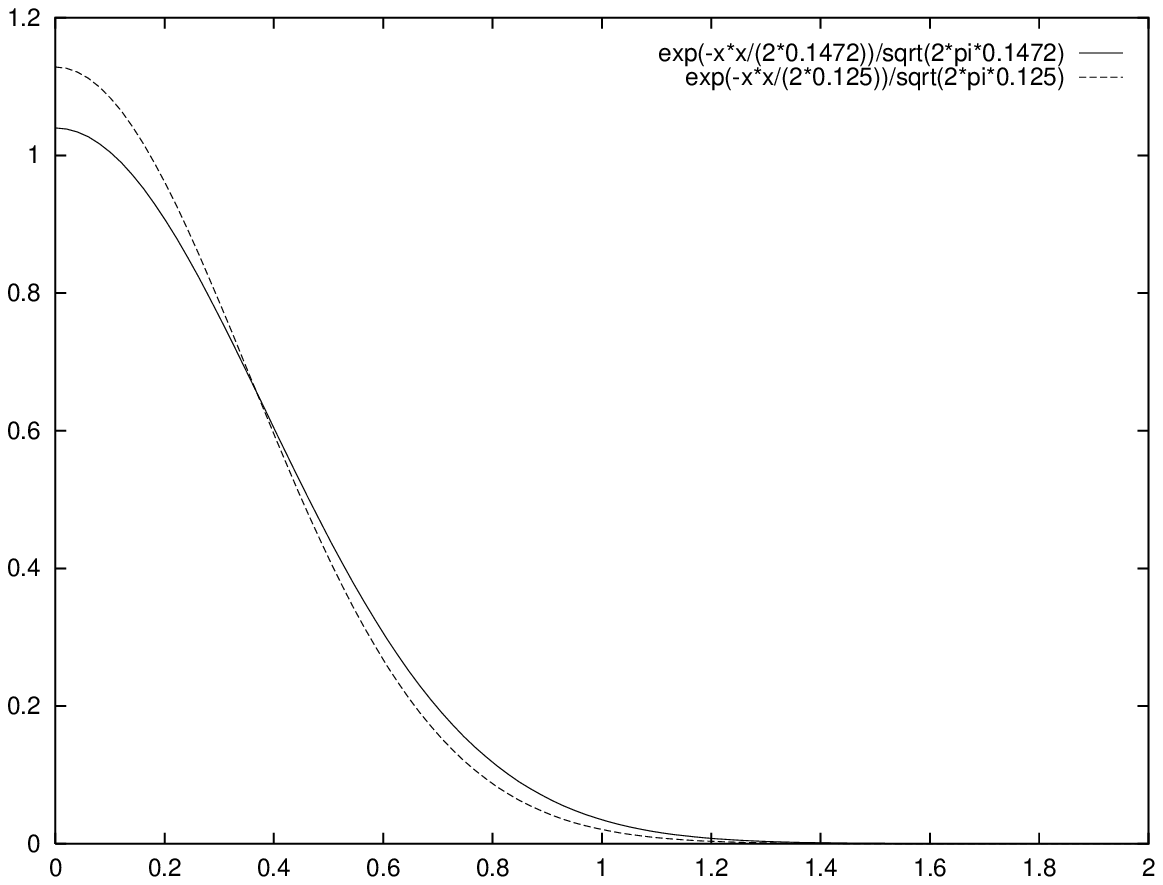}}
\centerline{}	
\vskip 20pt
\caption{The RG density of the anharmonic oscillator potential 
as compared to the variational density (dashed line).}
\end{figure}

\begin{figure}
	\epsfxsize=5cm
	\epsfysize=5cm
	\centerline{\epsffile{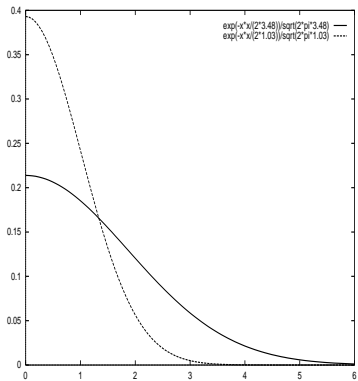}}
\centerline{}	
\vskip 20pt
\caption{Comparison of the particle density of the double well 
potential between the RG density and the variational density (dashed line). 
The RG density is globaly much better than the variational one (to 
compare with exact result see \protect\cite {kleinert}) but our 
approximation is not good enough to show the double picks structure of the 
true density (see text).}
\end{figure}

\end{document}